\documentclass[%
 reprint,
]{revtex4-2}
\usepackage{graphicx}
\usepackage{dcolumn}
\usepackage{bm}
\usepackage[dvipsnames]{xcolor}
\usepackage{siunitx}
\usepackage{mathtools}
\usepackage{svg}
\usepackage{amsmath}

\begin{document}

\preprint{APS/123-QED}

\title{Direct Observation of Sub-Poissonian Temporal Statistics in a Continuous Free Electron Beam with Sub-picosecond Resolution}

\author {S. Borrelli} 
\email{s.borrelli@tue.nl} 
\author {T.C.H. de Raadt } 
\author { S. B. van der Geer } 
\author {P.H.A. Mutsaers} 
\author {K.A.H. van Leeuwen} 
\author  {O.J. Luiten}
\affiliation{%
Department of Applied Physics, Eindhoven University of Technology, Groene Loper 19, 5612 AP, Eindhoven, The Netherlands}

\date{\today}

\begin{abstract}
We present a novel method to measure the arrival time statistics of continuous electron beams with sub-ps resolution, based on the combination of an RF deflection cavity and fast single electron imaging. 
We observe Poissonian statistics within time bins from 100~ns to 2~ns and increasingly pronounced sub-Poissonian statistics as the time bin decreases from 2~ps to 340~fs. This 2D streak-camera in principle enables femtosecond-level arrival time measurements, paving the way to observing Pauli blocking effects in electron beams and thus serving as an essential diagnostic tool towards degenerate electron beam sources for free electron quantum optics. 
\end{abstract}

\maketitle
The concept of anti-bunched electron beams exhibiting sub-Poissonian statistics has emerged as a topic of great significance in electron microscopy and lithography, as well as in the rapidly expanding field of free electron quantum optics. 
Sub-Poissonian electron beams hold immense potential as shot-noise-reduced probes for electron imaging with ultra-high resolution and enhanced signal-to-noise ratio \cite{kolobov2007quantum,mandel1995optical,magana2019quantum}. Furthermore, they are essential for establishing advanced quantum imaging techniques like ghost imaging and quantum holography \cite{pittman1995optical, abouraddy2001quantum,jack2009holographic}.
Investigating the electron arrival time statistics and correlations plays a pivotal role in two aspects: understanding the fundamental behavior of free electrons and designing sub-Poissonian electron sources.
The pioneering works of Kiesel et al. \cite{kiesel2002observation} and Kodama et al. \cite{kodama2011correlation} report the observation of anti-correlation between free electrons in coincidence experiments within time windows of 26~ps and 200~ps, respectively. 
More recently, Meier et al. \cite{meier2023few} and Haindl et al. \cite{haindl2023coulomb} observed the presence of strong energy anti-correlation between a few electrons confined in photoemitted pulses from nanometric needle tips. Furthermore, Keramati et al. \cite{keramati2021non} and Kuwahara et al. \cite{kuwahara2021intensity} demonstrated anti-bunching effects between electrons generated by, respectively, a photoemission gun and a spin-polarized source, in coincidence counting measurements with a resolution up to 48~ps. 
However, the direct observation of sub-Poissonian statistics in continuous electron beams within time windows of a few hundred fs has remained a challenging endeavor.

In this letter, we present a novel method for quantifying the statistical properties of a continuous electron beam with sub-picosecond resolution. 
We provide experimental results demonstrating the proposed technique's capability to continuously measure the electrons' concurrent arrival on a detector within time windows as short as a few hundred femtoseconds. 
These achievements are made possible through the unique combination of microwave-cavity-based electron beam deflection into a transverse Lissajous pattern followed by fast event-based electron imaging. Leveraging this combination, we have successfully developed a two-dimensional streak camera capable of sub-ps resolution.
To the best of our knowledge, the proposed method enabled for the first time the direct observation of sub-Poissonian statistics of a continuous electron beam across time scales ranging from picoseconds to hundreds of femtoseconds.
Our finding revealed that while electrons are randomly distributed according to Poisson statistics over time windows from 100 nanoseconds to 1 nanosecond, the emergence of anti-bunching effects becomes apparent at picosecond time scales. 

Measuring the electron arrival time statistics involves counting the number of electrons impinging on a detector within a given time window and determining the corresponding statistical distribution. One major limitation in such studies is the temporal resolution of the detector used to record electron events.
Despite impressive progress, the achievable temporal resolution is still in the order of a few tens of picoseconds \cite{gys2015micro,tremsin2020overview,carnesecchi2019performance,powolny2011time,kempers2023photodiode}.
Before applying the proposed method to explore the sub-ps time scale, we conducted benchmark studies to showcase the achievable results relying only on state-of-the-art detectors.

\begin{figure}
\includegraphics[clip,trim=0 5 5 5, width=0.95\columnwidth]{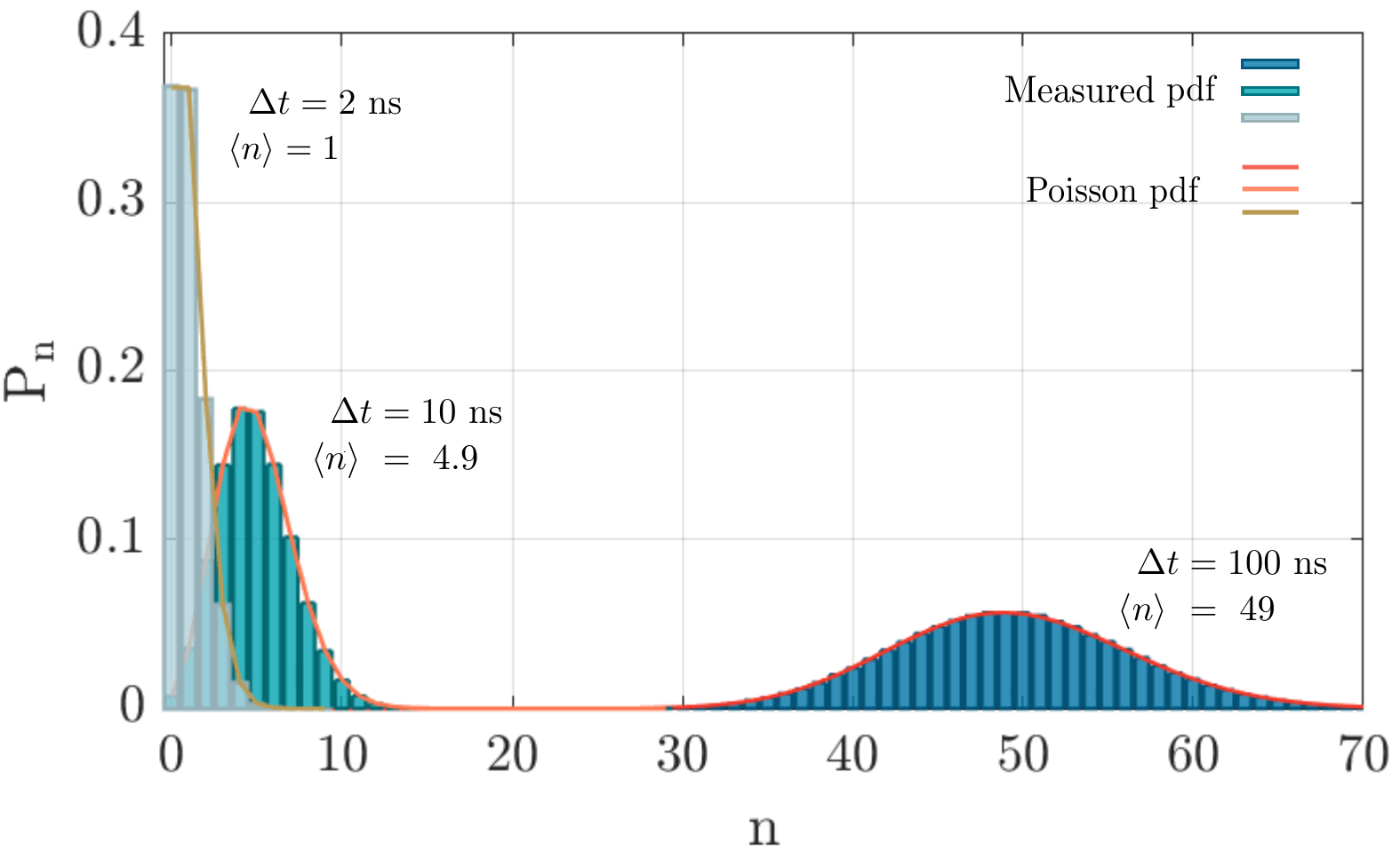}
\caption{\label{fig:Poisson_cheetah_resolution} Distribution of the simultaneous electron occurrences measured in $\Delta t = 100, 10, 2$~ns using the Timepix3 camera and the corresponding expected Poisson distribution.
Total number of analyzed electron events is $\simeq26$M for $\Delta t = 100, 10, 2$~ns.}
\end{figure}

In the ultrafast transmission electron microscope (UTEM) at Eindhoven University of Technology (TU/e), we generated a 200~keV continuous electron beam at a current $I~\sim~0.1$~nA and used a Timepix3 direct electron camera \cite{witten2001} to record electron events within a specific time window $\Delta t$. 
Figure~\ref{fig:Poisson_cheetah_resolution} presents the probability distributions $P_\mathrm{n}$ of the measured simultaneous occurrences of $n = 0, 1,...$ electrons in the selected time bins $\Delta t = 100~\mathrm{ns}, 10~\mathrm{ns}, 2~\mathrm{ns}$. 
Electron emission from a Schottky field emission gun can be modeled as a Poisson distribution, assuming a constant emission rate over time. 
The probability of $n$ electrons being emitted in $\Delta t$ is given by $P_\mathrm{n} = \frac{\langle n \rangle^n e^{-\langle n \rangle}}{n!}$, where $\langle n \rangle = \frac{\langle I \rangle}{e}\, \Delta t$. 
The solid curves in Fig.~\ref{fig:Poisson_cheetah_resolution} represent the expected Poisson probability density function for $\langle n \rangle$ equal to the average number of electrons we measured in each $\Delta t$.
Clearly, the measured distributions are well described by a Poisson distribution, with a residual sum of squares of the order of $10^{-6}$ in all three cases. 

The Timepix3 temporal resolution of 1.56~ns (per pixel) limits the study of the electron beam statistics to $\Delta t = 2$~ns at best. 
At shorter time scales, a loss of the random nature in the electron arrival time distribution may be observed due to the emergence of electron-electron anti-correlations. 
On one side, these correlations manifest the quantum-mechanical fermionic nature of electrons. According to the Pauli exclusion principle, the presence of an electron in a particular state inhibits the emission of another electron in the same state ({\it Pauli blockade}).
Consequently, the probability of two or more electron emissions within the coherence time of the beam is reduced, resulting in a sub-Poissonian distribution \cite{kuwahara2021intensity,silverman1987feasibility}. On the other side, the Coulomb repulsion between electrons emitted from the source's tip can also influence the dynamics and induce electron-electron anti-correlation within the characteristic time window of the interaction \cite{keramati2021non, kodama2019mechanism}. The method we propose here enables the investigation of time scales where the effects of Coulomb repulsion or the Pauli blockade become relevant in shaping the statistical distribution of electrons.
 
\begin{figure}
\includegraphics[clip,trim=0 30 40 0, width=0.95\columnwidth]{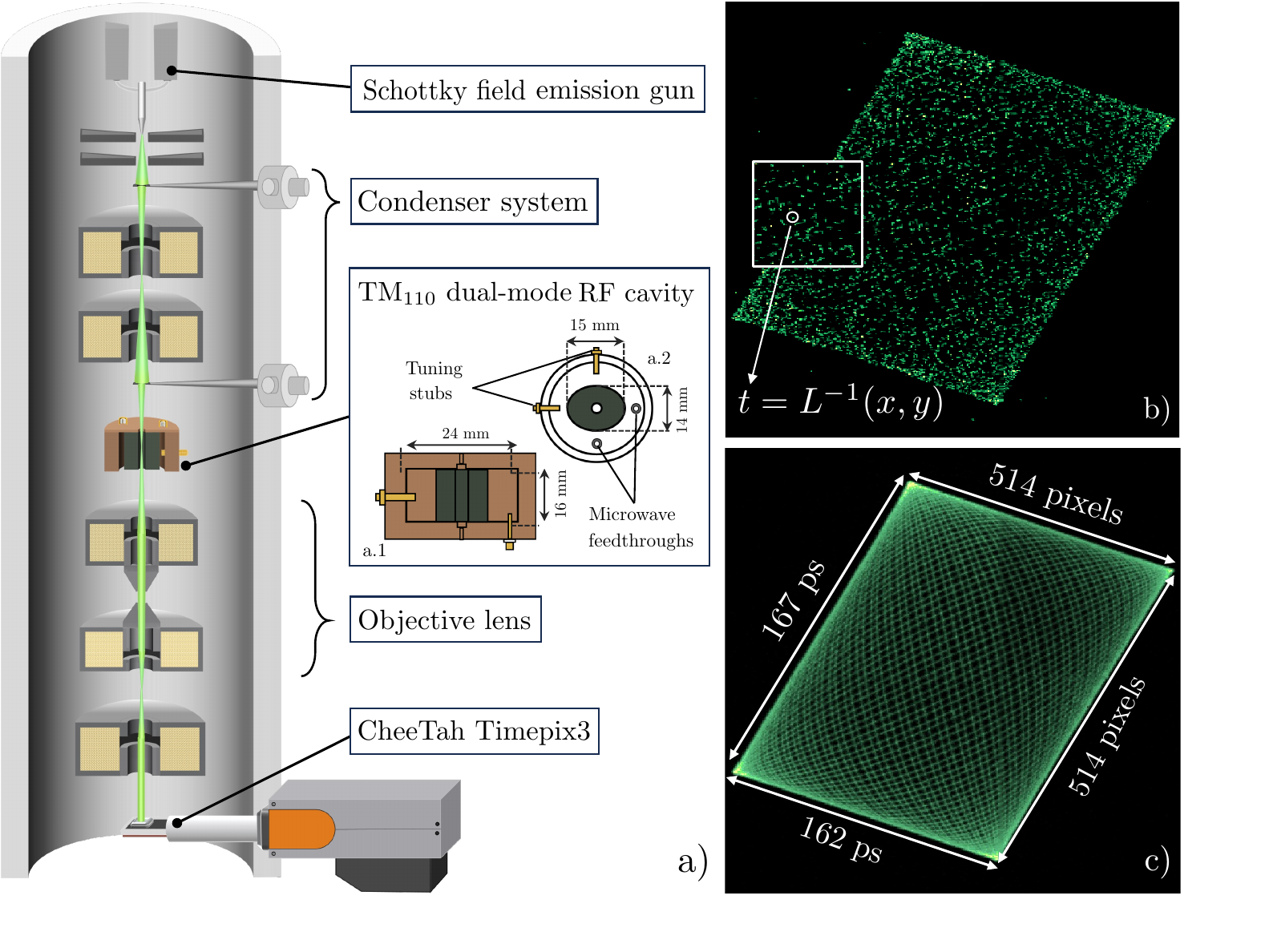}
\caption{\label{fig:Fig2}  a) Illustration of the RF-cavity-based UTEM at TU/e. In the inset, schematic representations include a (a.1) cut-through view and (a.2) top view of the RF cavity. b) Lissajous pattern at $I = 0.1~nA$, imaged on the Timepix3  with \SI{5}{\micro\s} exposure. c) Sum of 200 images.}
\end{figure}

Figure \ref{fig:Fig2}a presents the layout of the TU/e RF-cavity-based ultrafast transmission electron microscope \cite{Borrelli2023} (UTEM).  
This UTEM employs a commercial 200~keV FEI Tecnai~TF20, generating a partially coherent continuous free electron beam with 1~eV energy spread. The microscope is equipped with an RF cavity operated in TM$_{110}$ dual mode at resonance frequencies of 3.000~GHz and 3.075~GHz, derived from the same 75~MHz driving signal \cite{van2018dual}. 
The resulting electromagnetic field configuration induces a two-dimensional periodic deflection of the continuous electron beam into a transverse Lissajous pattern (see Fig.~\ref{fig:Fig2}c) consistently generated at the same repetition rate as the driving signal, corresponding to a period of $13.3$~ns.
This pattern is detected using a Timepix3 direct electron camera mounted in the fluorescent screen chamber of the microscope \cite{witten2001}. 
This hybrid pixelated detector comprises a silicon sensor with an array of $514 \times 514$ pixel detectors. It allows for independent measurements of energy deposition (Time-over-Threshold, ToT) and timing information (Time-of-Arrival, ToA) of single electrons, as well as the determination of the position of the electron-activated pixels on the detector \cite{poikela2014timepix3}. 
Upon striking the detector, an electron triggers the activation of a cluster of pixels featuring a given size, total ToT, and ToA values. 
Through a comprehensive characterization of the Timpepix3 response to 200~keV single electrons, we determined the average number of pixels involved in a single electron hit (7 pixels) and the maximum variation in ToA values among the pixels within a cluster (30~ns). 
Additionally, we analyzed the distribution of the cumulative ToT across the pixels within a cluster and determined the average ToT of a single electron hit (7000~arbitrary units). 
This characterization enabled the development of a clustering algorithm that allowed us to reconstruct the timing and position of individual electrons incident on the detector from the activated cluster of pixels \cite{SuppMaterial}. 
 
Figure \ref{fig:Fig2}b shows a Lissajous pattern generated in the RF-cavity-based UTEM at $I = 0.1$~nA and imaged on the Timepix3 detector with \SI{5}{\micro\s} exposure.
Each green dot in Fig.~\ref{fig:Fig2}b represents a single electron that has landed on the detector at a specific time instant. The position of each electron-dot is uniquely identified by the ($x,y$) coordinate on the detector, assigned through the developed clustering algorithm. 
At the same time, the arrival time of each electron is recorded with nanosecond resolution.
The Lissajous pattern's features are clearly recognizable in Fig.~\ref{fig:Fig2}c, which displays a composite pattern formed by all the electrons accumulated in 75000 Lissajous periods.
Mathematically, this pattern is described by the vectorial function $ L: t \mapsto (x, y)$ ({\it Lissajous function}) defined as \cite{SuppMaterial}
\begin{align*}
    x(t) &= A_1\left[ \sin\left(\alpha_1+  K_1 \right) - \sin(\alpha_1) - K_1 \cos(\alpha_1)\right] \\
     y(t) &=  A_2\left[ \sin\left(\alpha_2 +  K_2\right) - \sin(\alpha_2) -  K_2 \cos(\alpha_2)\right], \nonumber
\end{align*}
where $x(t)$ and $y(t)$ represent the transverse coordinates at the cavity exit of a generic electron in the beam along the horizontal and vertical axis of a Cartesian coordinate system. 
In these equations, $K_{1,2}\equiv~\frac{L_{\mathrm{cav} \,} \omega_{1,2}}{\mathrm{v}_\mathrm{z}}$ with $L_\mathrm{cav}$ being the cavity length, $\omega_1~=~2\pi\times~3.000$~GHz and $\omega_2~=~2\pi\times3.075$~GHz the two cavity's resonance frequencies, and v$_\mathrm{z}$ the electron velocity; $\alpha_{1,2} (t)~\equiv~\omega_{1,2} t +\phi_{1,2}$ with $\phi_{1,2}$ denoting the phases of the cavity's fields at $t = 0$; $A_{1,2}~\equiv~+(-)\frac{q_\mathrm{e} \mathrm{v}_\mathrm{z}}{m \gamma} \frac{B_{1,2}}{\omega_{1,2}^2}$ with $q_\mathrm{e}$ being the electron charge and $\gamma$ the Lorentz factor.
The amplitude factors $A_{1,2}$ measure the size of the $x,y$ side of the Lissajous pattern and depend on the RF power $W_{1,2} \propto B_{1,2}^2 $ feeding the corresponding cavity mode.
The phase difference between the two cavity modes $\phi_1 - \phi_2$ determines the spacing of the inner lines within the Lissajous figure. 

\begin{figure*}
\includegraphics[clip,trim=0 0 0 0, width=0.95\textwidth]{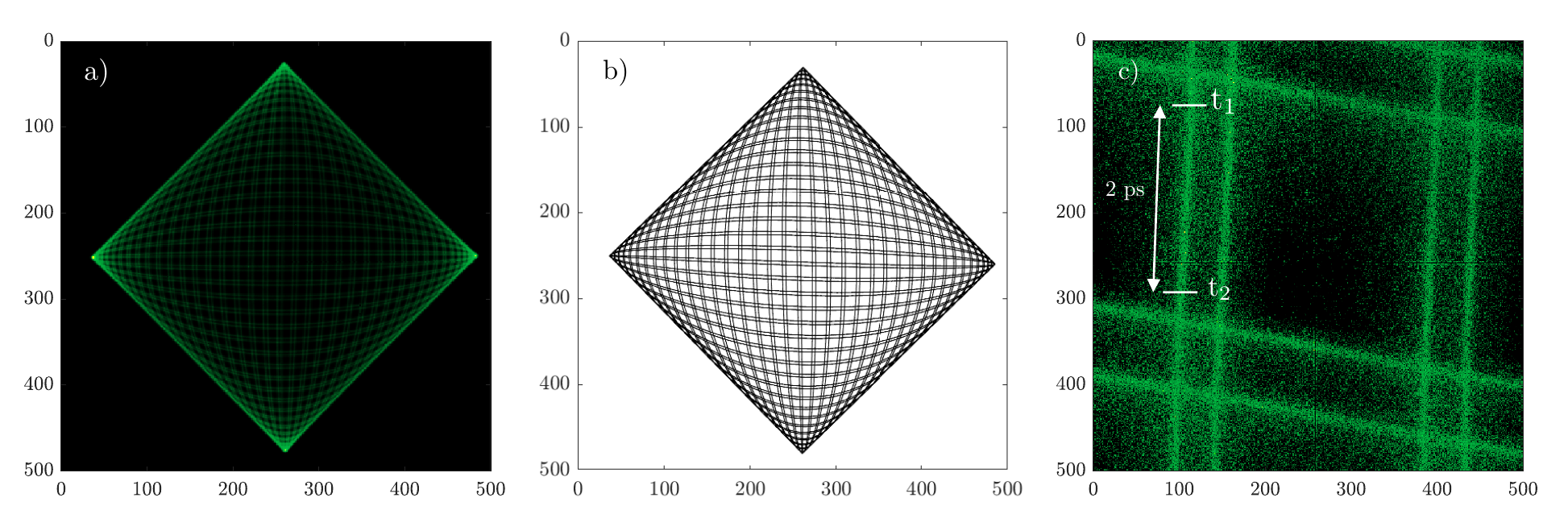}
\caption{\label{fig:Fig3} a) Composite image showing the sum of 1000 Lissajous patterns at $I = 0.1$~nA. b) Lissajous function $L(t)$ fitted to the measured pattern \cite{SuppMaterial}. c) Composite image of six overlapped zoomed-in Lissajous patterns at $I = 46.2$~nA, each magnified 1850 times and captured using the Timepix3 detector with a \SI{50}{\micro\s} exposure.}
\end{figure*}

One noteworthy characteristic of the Lissajous pattern is its time-dependent nature, associating a precise time instant with each position within the pattern.
When considering a generic electron in a Timepix3-captured Lissajous pattern, the nanosecond-resolution ToA measurement provides information about the Lissajous period during which the electron landed on the detector within the exposure time. 
Additionally, the measured electron position (x,y) within the pattern enables precise determination of its arrival time within the $13.3$~ns Lissajous period. 
When imaging the entire Lissajous pattern across the $514 \times 514$ pixels of the detector, a maximum resolution of 315~fs per pixel can, in principle, be attained. However, zooming in on the Lissajous figure using the microscope's imaging system allows for capturing only a fraction of the pattern corresponding to 1-2 picoseconds. 
In this configuration, a resolution of a few femtoseconds per pixel can be achieved. 

We employed this technique to study the arrival time statistics of a continuous free electron beam generated in our UTEM at $I = 46.2$~nA. 
During the experiment, the beam was transversally deflected into a Lissajous pattern and imaged on the Timepix3, using the cavity's driving signal as a timestamp for the detector. 
For the data analysis, we utilized a fitting algorithm \cite{SuppMaterial} to fit the Lissajous function $L(t)$ to a measured pattern (see Fig.~\ref{fig:Fig3}a and \ref{fig:Fig3}b). 
To enhance our temporal resolution, we conducted a statistical analysis on the 1850 times magnified Lissajous pattern in Fig.~\ref{fig:Fig3}c, captured with an exposure time of \SI{50}{\micro\s}. The total beam current on the detector in this magnification mode is $I_{\mathrm{det}} = 0.12$~nA.
This magnification yields a temporal resolution of approximately $\sim30$~fs per pixel when other sources of uncertainty are neglected.
A careful microscope alignment enabled magnifying around a central point without any rotation or shifting of the pattern. 
Consequently, the same Lissajous function evaluated in the best-fit parameters accurately describes the magnified pattern once enlarged by the same magnification factor.  
We increased the number of acquired events by continuously repeating the imaging process 1408 times, achieving a total exposure of \SI{18.03}{s}.

To investigate the electron arrival time statistics, we selected the specific time bin $\Delta t$ defined by the line segment between the points $t_1$ and $t_2$ in Fig.~\ref{fig:Fig3}c. Given the measured coordinates of these two points, we determined the corresponding times within a Lissajous period through the inversion of the Lissajous function evaluated in the best-fit parameters. Accordingly, the selected line segment corresponds to $\Delta t = 2$~ps. 
In selecting the time bin, we excluded the crossing points between two inner lines and considered only the portion of the Lissajous pattern where the function is invertible \cite{SuppMaterial}.
In this case, for each electron event within the selected time bin, we could establish a Lissajous period based on the ToA measurement and determine the precise arrival time within the period by inverting the fitted Lissajous function. In practice, we tallied the number of electrons arriving at the detector within $\Delta t = 2$~ps for each measured Lissajous period. At later stages, we performed a more detailed analysis by partitioning the line segment between the points $t_1$ and $t_2$ in Fig.~\ref{fig:Fig3}c into finer divisions of 1~ps, 500~fs, and 340~fs.

\begin{figure*}
\includegraphics[clip,trim=10 10 15 5, width=0.97\textwidth]{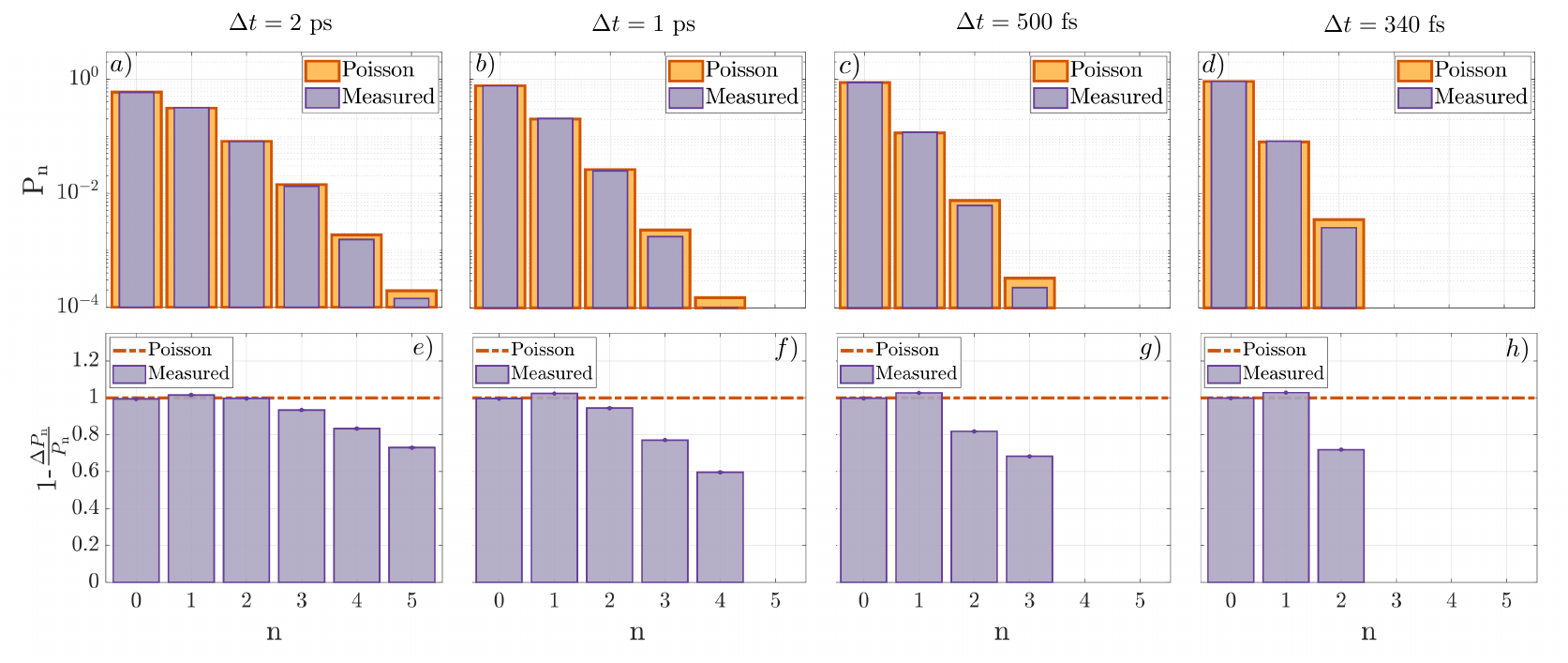}
\caption{\label{fig:46nA} Figures (a) to (d) show the distributions of the measured number of 0, 1, 2, 3, 4, and 5 simultaneous electron events, alongside the corresponding expected Poisson distributions with  $\langle n \rangle = 0.52~\mathrm(a),~0.26~\mathrm(b),~0.13~\mathrm(c),~0.09~\mathrm(d)$, for different time windows $\Delta t$. 
The corresponding deviations from the expected Poisson distribution are shown in (e) to (h). 
The dotted orange line represents the expectation for randomly distributed data according to a Poisson distribution. 
} 
\end{figure*}

Figures~\ref{fig:46nA}a-d present the measured probability $P_\mathrm{n}$ of 0,1,2,3,4 and 5 simultaneous electron events in the selected time bins $\Delta t~=~2~\mathrm{ps},~ 1~\mathrm{ps},~ 500~\mathrm{fs},~ 340~\mathrm{fs}$, shown in purple bars. The orange bars represent the expected Poisson distribution $\Tilde{P}_\mathrm{n} = \frac{\langle n \rangle^n e^{-\langle n \rangle}}{n!}$, with $\langle n \rangle$ equal to the measured average number of electrons in each $\Delta t$. 
The measured number of 0, 2, 3, 4, and 5 electron events is found to be smaller than expected from a Poisson distribution, while the number of single electron events is larger than predicted by a Poissonian behavior.
These observed deviations are the signature of a sub-Poissonian distribution.
The measured discrepancy is further evident in the plots in Fig.~\ref{fig:46nA}e-h, displaying for every $\Delta t$ the relative deviation of the measured distribution from the expected Poisson distribution, expressed as $1 - \frac{\Delta P_\mathrm{n}}{\Tilde{P}_\mathrm{n}}$, with $\Delta P_\mathrm{n} = P_\mathrm{n} - \Tilde{P}_\mathrm{n}$. 
The displayed error bars, though challenging to discern, represent the statistical uncertainties associated with observing $N_\mathrm{n}$ simultaneous $n$ electron events, computed as $ \frac{\sqrt{N_\mathrm{n} \left(1- \frac{N_\mathrm{n}}{N_{\mathrm{exp}}} \right)}}{\widetilde{P}_\mathrm{n} \;  N_{\mathrm{exp}}}$, where $N_{\mathrm{exp}}$ is the total number of measurements. 
Considering the measurements for $\Delta t = 2$~ps (Fig.~\ref{fig:46nA}a and \ref{fig:46nA}e), we observe a reduction of $0.07\%$, $0.1\%$, $7\%$, $17\%$, $26\%$ in the number of 0, 2, 3, 4, and 5 electron events ($8\times10^8$, $1.1\times10^8$, $1.8\times10^7$, $2.1\times10^6$,$2.1\times10^6$, $2.9\times 10^5$), respectively. In comparison, the number of single electron events ($4.2\times10^8$) shows an increase of $2\%$ compared to the expected Poisson distribution.
The suppression of multiple electron events in favor of single electron events becomes even more pronounced when reducing the observation time window. Within the 340~fs time bin (Fig.~\ref{fig:46nA}d and \ref{fig:46nA}h), the occurrences of 0 and 2 electron events ($7.4\times10^9$ and $2\times10^7$) drop by $0.2\%$ and $28\%$, respectively. Meanwhile, the count of single electron events ($6.7\times10^8$) experiences a $3\%$ increase. 
It should be noted that as the time bin is reduced, the number of recorded events correspondingly decreases. 
Consequently, statistical errors for $n>2$ can become too high to draw significant conclusions when examining very short time scales. 
For this reason, at 340~fs, we only present the data relative to the simultaneous arrival of 2 electrons on the detector, while at longer time windows, we can still contemplate 2, 3, and even 5 electron events.  
The dataset included in this study amounts to a total of 1.3 billion registered events.

We have measured arrival time statistics of continuous free electron beams with $\sim 300$~fs resolution in a 200~keV UTEM.  
We find that the statistical distribution of electrons is Poissonian within time windows from 100~ns down to 1~ns, while we observe increasingly pronounced sub-Poissonian statistics as the counting time window decreases from 2~ps to 340~fs.
The temporal resolution in the measurements was limited by the hit rate on the detector exceeding the maximum hit rate of the Timepix3 ($1.2 \times 10^8~\mathrm{s}^{-1}$) at $I \sim $~few nA \cite{SuppMaterial}.
A higher resolution can be achieved by incorporating a fast beam blanker in the microscope column to reduce the effective hit rate on the detector.

Based on the current experimental results, it is challenging to ascertain unambiguously whether the observed anti-bunching is primarily caused by Coulomb repulsion among the electrons in the beam or quantum fermionic statistics. 
Properly discriminating between the two effects and establishing the corresponding time scales is an ongoing challenge \cite{kodama2019mechanism}.
Further experimental and theoretical investigations are thus necessary to explore the mechanism underlying the appearance of the two kinds of anti-bunching.
However, Pauli blocking effects are expected to become significant only at sub-fs timescales for an unpolarized electron beam with an energy spread of 1~eV \cite{SuppMaterial}.  
Additionally, we performed a classical charge particle tracing simulation of a 46.2~nA electron beam emission from a simplified yet realistic model of the Schottky field emission gun in our UTEM incorporating all pairwise stochastic Coulomb interactions in GPT \cite{Bas2022,van20053d,poplau2004multigrid}. Preliminary results reveal that sub-Poissonian behavior becomes evident at a distance less than $\sim1$~mm from the emission tip within time windows of 2~ps or shorter \cite{SuppMaterial}. 
Achieving a higher temporal resolution makes it possible to delve into electron beam statistics on even shorter time scales.
Furthermore, combining this advancement with a systematic study encompassing different electron beam extraction currents and possibly a spin-polarized electron source \cite{kuwahara2021intensity} would enable distinguishing Coulomb effects from Pauli blockade.  
While observing the Fermi-Dirac statistics holds significant importance from a fundamental physics perspective, gaining a deeper understanding of the dynamics underlying Coulomb repulsion between electrons in a beam is of great interest in free electron quantum optics. 
These insights pave the way for the development of advanced techniques for electron beam manipulation and the realization of quantum electron sources for ultra-high resolution interaction-free electron imaging \cite{kruit2016designs,boggild2017two}. 
Furthermore, the method proposed and demonstrated in this study has the potential to shed light on the possible correlation between the arrival time statistics of electrons and radiation damage \cite{vandenbussche2019reducing, kisielowski2019discovering}.
Finally,  the versatility of the presented method extends its applicability beyond the specific case studied here, making it suitable for investigating and comparing the statistical properties of various electron sources. \\

We are deeply grateful to Hein van den Heuvel and Harry van Doorn for their invaluable assistance in setting up the experiment and their prompt resolution of last-minute electronic challenges. 
Our heartfelt appreciation also goes to Stefan Kempers, Jim Franssen, and Erik Kieft for the enlightening discussions, constant support, and invaluable suggestions.

O.L. conceived the original idea. S.B., T.R., P.M., K.L., and O.L. collectively designed the experimental setup, established the analysis procedure, and explored the theoretical framework.
T.R. configured the Timepix3 detector for measurements and developed the data acquisition software. S.B. conducted preparatory simulations, analytically derived the fitting function, and prepared the microscope for the experiment. S.B. and T.R. jointly conducted the experiment, performed the data analysis, and generated the plots. S.B. and S.G. performed the simulative study in GPT. S.B. wrote the manuscript, incorporating valuable inputs from all co-authors.

\nocite{*}

\bibliography{1paper}

\end{document}


\preprint{APS/123-QED}

\title{Direct Observation of Sub-Poissonian Temporal Statistics in a Continuous Free Electron Beam with Sub-picosecond Resolution}

\author {S. Borrelli} 
\email{s.borrelli@tue.nl} 
\author {T.C.H. de Raadt } 
\author { S. B. van der Geer } 
\author {P.H.A. Mutsaers} 
\author {K.A.H. van Leeuwen} 
\author  {O.J. Luiten}
\affiliation{%
Department of Applied Physics, Eindhoven University of Technology, Groene Loper 19, 5612 AP, Eindhoven, The Netherlands}

\date{\today}
\maketitle
\onecolumngrid
\renewcommand\thesection{S\arabic{section}}
\renewcommand\thefigure{S\arabic{figure}}
\renewcommand\thetable{S\arabic{table}}

\section{Clustering algorithm for single electron events reconstruction in a Timepix3 measurement}
In this study, we recorded single electron events using a Timepix3  hybrid pixel detector (EM CheeTah T3, Amsterdam Scientific Instruments \cite{poikela2014timepix3, witten2001}) installed in the fluorescent screen chamber of our transmission electron microscope \cite{Borrelli2023}. This camera provides a continuous data stream, including information on the position of the pixels activated by electron impacts, the energy deposition (Time-over-Threshold, ToT) on each pixel, and the activation time (Time-of-Arrival, ToA) of each pixel with a temporal resolution of 1.56~ns.
Each pixel comprises an individual charge collector and analog-to-digital converter circuit. The Time-over-Threshold (ToT) is the number of nanoseconds the signal from the charge collector stays above a threshold value, which is related to the electron energy. The Time-of-Arrival (ToA) is the time at which the signal first surpasses this threshold.

The foundation of this work is rooted in the capacity to distinguish single electron impacts. 
In preparation for our experiment, a crucial undertaking thus involved probing the response of the Timepix3 camera to 200~keV single electrons. 
When an electron impinges on the active area of the detector, it triggers the activation of a cluster of pixels.
Through an in-depth statistical analysis, we identified key parameters 
encompassing the average size of the pixel cluster activated by a single electron event, the maximum variation in Time-of-Arrival (ToA) among the pixels engaged in such an event, and the distribution of the cumulative Time-over-Threshold (ToT) over the pixels within the cluster.
These parameters were subsequently harnessed within a clustering algorithm enabling the reconstruction of the time and position of each individual electron impinging on the detector from the activated pixels cluster.

\begin{figure*}
\includegraphics[clip,trim=0 0 0 0, width=0.95\textwidth]{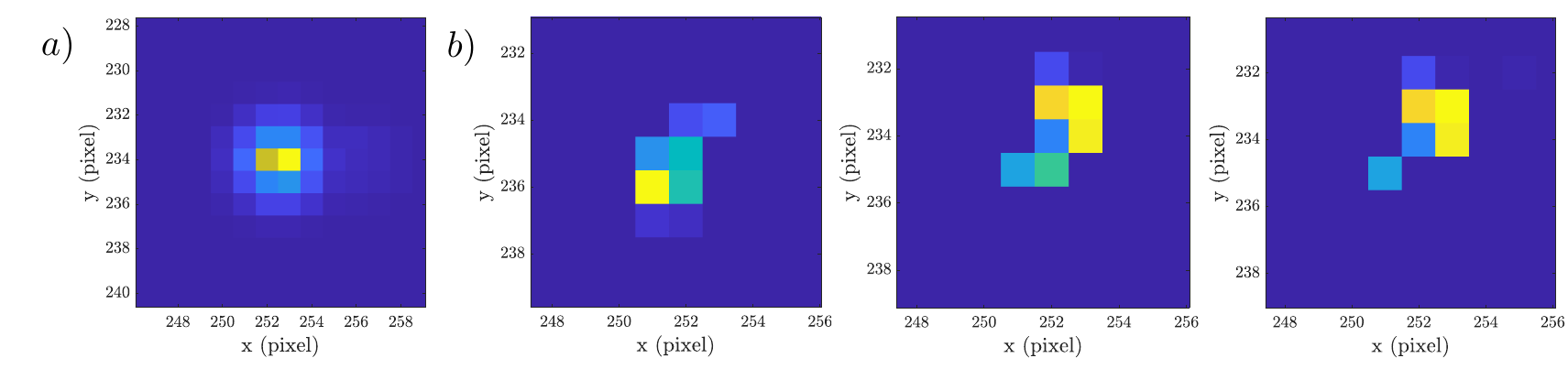}
\caption{\label{fig: Fig1_SuppMat} a) Sum of 100 images of a pulsed electron beam at $I_{\mathrm{cont}}= 0.098$~nA, each captured using the Timepix3 detector with a \SI{1}{\s} exposure. b) Collection of images of single electron pulses on the Timepix3 detector.}
\end{figure*}

To carry out this characterization, we operated the microscope in pulsed mode \cite{verhoeven2018high} at a  continuous current $I_{\mathrm{cont}}~=~0.098$~nA, measured using a separate detector from the Timepix3. The setup generated electron pulses with duration $\tau \approx 250$~fs at a repetition rate $\nu = 75$~MHz. Employing the Timepix3 detector, we captured a series of 100 images of the pulsed beam, each with an exposure time $T_{\mathrm{exp}} = 1$~s. The sum of these images is presented in Figure~\ref{fig: Fig1_SuppMat}a. 
Since each exposure encompasses multiple electron pulses, to characterize the response of the Timepix3 to individual electrons, we divided $T_{\mathrm{exp}}$ into sub-intervals of 100~ns each. 
At a continuous current $I_{\mathrm{cont}}$, the pulsed current corresponding to a pulse length $\tau \approx 250$~fs is approximately $I_{\mathrm{pulse}} \approx 1.8$~fA ($I_{\mathrm{pulse}} = I_{\mathrm{cont}} \; \tau \, \nu$). Consequently, the expected average count of electrons within the $\Delta t = 100$~ns time frame is $N = \frac{I_{\mathrm{pulse}} \; \Delta t}{q_{\mathrm{e}}} = 0.012$. It is then reasonable to assume that, on average, only one electron hits the detector during the interval $\Delta t$. 
Figure~\ref{fig: Fig1_SuppMat}b presents a collection of images of single electron pulses, where the cluster of activated pixels is recognizable. The total number of single electron pulses identified is $\approx 1.1 \times 10^6$, which results in a measured pulsed current of 1.76~fA, in good agreement with our expectation. 

\begin{figure*}
\includegraphics[clip,trim=0 0 0 0, width=0.95\textwidth]{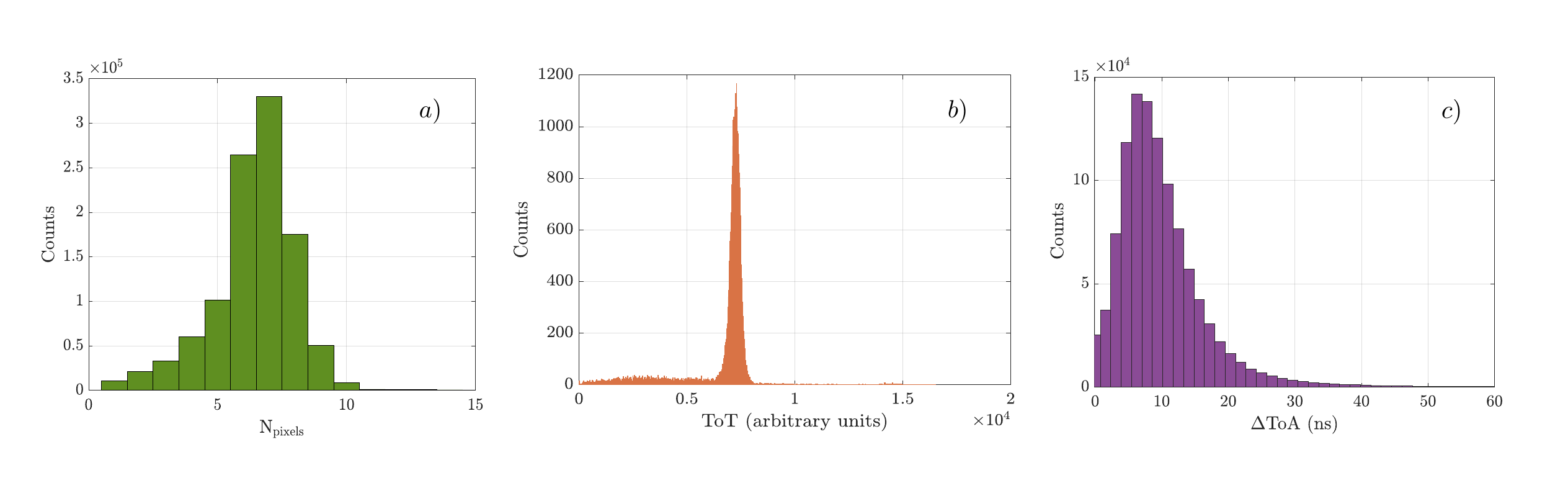}
\caption{\label{fig: Fig2_SuppMat} a) Distribution of the number of pixels activated in a single electron hit b) Distribution of the cumulative ToT, in arbitrary units, across the pixels engaged in a single electron hit. c) Distribution of the time difference $\Delta (\mathrm{ToA})$ between the maximum and the minimum ToA values among the pixels engaged in a single electron hit. All distributions are generated from a dataset of approximately $1.1 \times 10^6$ recorded single electron pulses. Each of these pulses was captured on the bottom left quadrant of the Timepix3 camera with an exposure of 1~s.}
\end{figure*}

From this series of single electron pulses, we derived the distribution of the number of pixels activated in a single electron hit, as illustrated in Fig.~\ref{fig: Fig2_SuppMat}a. 
This distribution yields an average pixel count of 7 per individual electron event. Additionally, we assessed the average maximum distance $\bar d$ between the activated pixels across all the recorded single electron events, which turned out to be about 5 pixels. 
Furthermore, we analyzed the distribution of the cumulative ToT value over the pixels engaged in a single electron hit, as shown in Fig.~\ref{fig: Fig2_SuppMat}b. The ToT value corresponding to the distribution mean represents the average cumulative ToT over the pixels within a single cluster, which is proportional to the single electron energy. This value amounted to $7240$~(arbitrary units), while the distribution variance is 210~(arbitrary units).
The histogram in Fig.~\ref{fig: Fig2_SuppMat}b exhibits two noteworthy characteristics. Firstly, it features a low plateau at ToT values between 0 and $\approx 0.6$ (arbitrary units). These counts at small ToT values are attributed to scattered electrons within the microscope column or background radiation captured by the detector.  
Secondly, a small peak appears around a ToT value of $\approx 14000$~(arbitrary units), approximately double the average cumulative ToT.
This peak corresponds to off-chance instances where two electrons arrive simultaneously, implying that certain pulses contain two electrons. 
Finally, we computed the distribution of the difference, $\Delta (\mathrm{ToA})$, between the maximum and minimum ToA values among the pixels within a cluster. The distribution, presented in Fig.~\ref{fig: Fig2_SuppMat}c, demonstrates an average $\Delta (\mathrm{ToA})$ of 6.25~ns. A reasonable upper limit to $\Delta (\mathrm{ToA})$ is $\approx 30$~ns. 
Interestingly, Ref~\cite{haindl2023coulomb} discusses a similar characterization of the Timepix3 at the same beam energy and shows comparable parameters.

\begin{table}
    \renewcommand{\arraystretch}{2.0} 
    \begin{tabular}{| c|c|c|c|c|c|c|c|c|}
        \hline 
        &
        \multicolumn{2}{|c|}{Top Right Quad} &
        \multicolumn{2}{|c|}{Bottom Right Quad}&
        \multicolumn{2}{|c|}{Bottom Left Quad} &
        \multicolumn{2}{|c|}{Top Left Quad} \\
        \hline
        Exposure (s) & 1  & 0.1  & 1  & 0.1  & 1  & 0.1  & 1  & 0.1  \\
        \hline
        Average ToT (a.u.) & 6564  & 6555  & 6820  & 6861  & 7090  & 7077
        & 5397  & 5381  \\
         \hline
          Average $\sigma_{\mathrm{ToT}}$ (a.u.) &  201 &  198 &  229 &  223 &  239 &  233 & 204 & 201 \\ 
         \hline
        Average $N_{\mathrm{pixels}}$ & 7 & 7 & 6 &6 & 7 & 7& 7 & 7  \\
        \hline
        $\bar d$ (pixel) & 5 & 5& 5 & 5& 5 & 5& 5 & 5 \\[5pt]
        \hline
        Average $\Delta (\mathrm{ToA})$ (ns) & 4.7 & 4.7 & 6.2 & 6.2 & 6.2 &6.2 & 4.7 & 4.7 \\
        \hline
    \end{tabular}
    \label{tab:tab1}
    \caption{Average number of pixels activated in a single electron hit, average maximum distance $\bar d$ between the activated pixels, average spread in ToA $\Delta \mathrm{ToA}$, and average ToT and ToT variance, in arbitrary units (a.u.), among the pixels in the cluster. These parameters are measured for all four quadrants of the Timepix3 detector.}
\end{table}

The Timepix3 detector comprises four quadrants (chips), collectively forming a grid of $514 \times 514$ pixels. 
These quadrants are manufactured individually, leading to slight variations in their responses to incident electrons.
To account for these manufacturing deviations, we conducted a chip-specific characterization.
The process involved directing single electron pulses to an individual chip and performing the analysis described above for each chip at two different exposure times (1 s and 0.1 s) to ensure that the detector was not overloaded. A summary of the outcomes is provided in Table~\ref{tab:tab1}. The distributions shown in Fig.~S2 pertain to the bottom left quadrant.

The characteristic parameters we have identified, resulting from a statistical analysis of the detector's response, constitute the essential components of our clustering algorithm.
The algorithm takes as input the Timepix3 output data stream, which contains the energy (ToT) and activation time (ToA) values for each triggered pixel during the measurement, along with the coordinates of the activated pixels. 
It groups pixels within the same cluster if their activation occurs within the upper limit of $\Delta(\mathrm{ToA})$, if they are within a distance of $\bar d$, and if the cumulative ToT across these pixels falls within 4$\sigma$ from the average cumulative ToT. 
The algorithm assigns the weighted cluster centroid, with ToT values serving as weights, as the position of the identified single electron event, and the cumulative ToT values among the pixels in the cluster as the event energy. 
The arrival time of the single electron event is determined as the weighted average of ToA over the pixels in a cluster, using the ToT values as weights. Thus, the ToA assigned to the i-th electron is calculated as ToA(i) = $ \frac{1}{\sum_{n_\mathrm{i}} ToT(n_\mathrm{i})} \sum_{n_\mathrm{i}} ToA(n_\mathrm{i}) *ToT(n_\mathrm{i})$, where $n_\mathrm{i}$ labels the pixels involved in the i-th electron hit. 
The uncertainty associated with the weighted ToA average for the i-th electron is given by $\sigma_{\overline{ToA}(i)} = \sigma \sqrt{\sum_{n_\mathrm{i}} w(n_\mathrm{i})^2}$, where $w(n_\mathrm{i}) = \frac{ToT(n_\mathrm{i})}{\sum_{n_\mathrm{i}} ToT(n_\mathrm{i})}$ and $\sigma$ represents the uncertainty in the ToA readings, set at 1.56~ns. Assessing this uncertainty results in $\sigma_{\overline{ToA}(i)} \leq 1.56$ consistently. 

\begin{figure}[]
\centering
\floatbox[{\capbeside\thisfloatsetup{capbesideposition={left, center},capbesidewidth=5cm}}]{figure}[\FBwidth]{\caption{Distribution of the cumulative ToT across the pixels involved in a single electron hit. This distribution was recorded at 46.2~nA beam current with a \SI{50}{\micro\s} exposure, exceeding the nominal maximum hit rate of the detector.}
\label{fig: Fig3_SuppMat} }%
{\includegraphics[clip,trim=50 30 25 80, width=7cm]{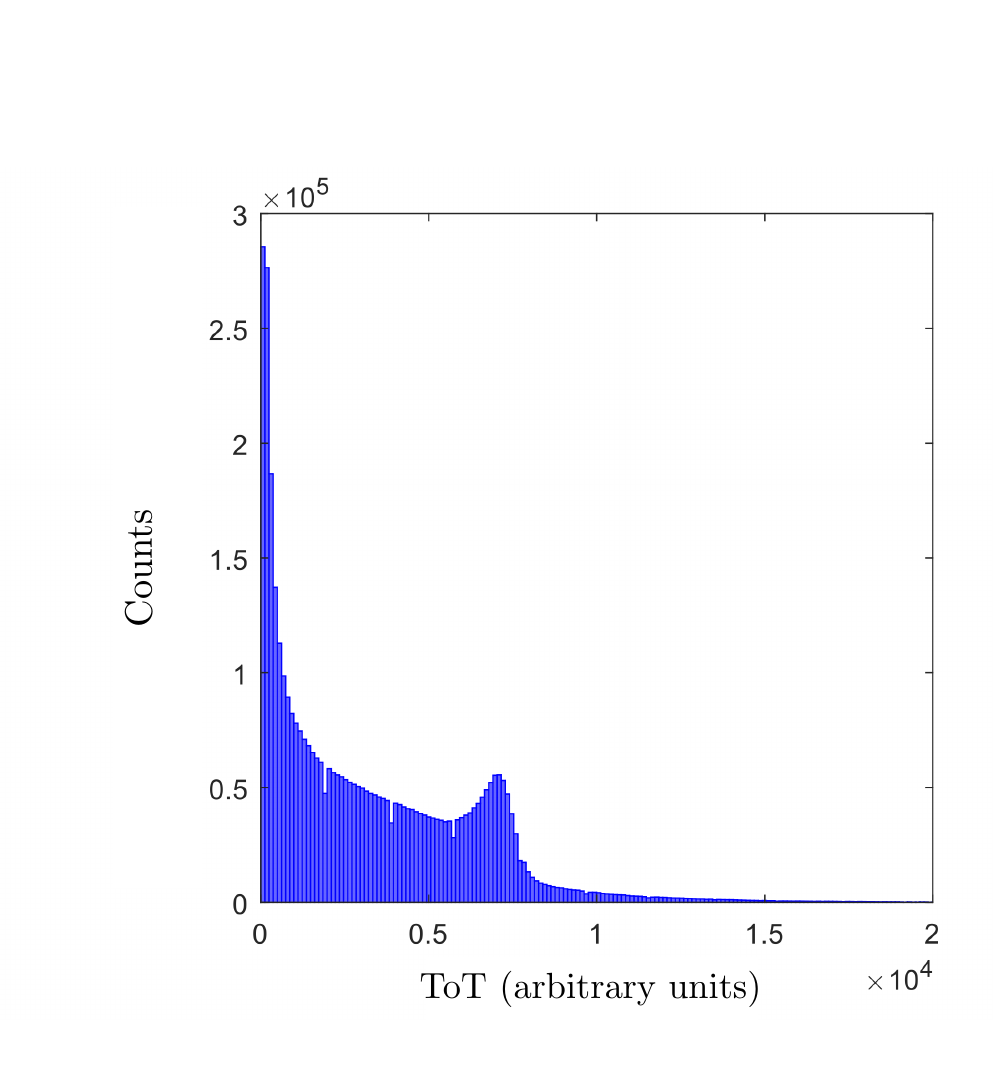}}
\end{figure}

It is important to highlight that when the Timepix3 is employed to measure electron currents surpassing a few hundred pA, the hit rate on the detector exceeds the nominal maximum allowed hit rate of $1.2 \times 10^8~\mathrm{s}^{-1}$. The data presented in Fig.~4 of the main paper were acquired at a beam current on the detector of $I = 0.12$~nA, resulting in a hit rate $\frac{N_{\mathrm{events}}}{T_{\mathrm{exp}}} \approx 7 \times 10^8$, where $N_{\mathrm{events}} = 13 \times 10^9$ electron events and $T_{\mathrm{exp}} = 18.03$~s.
This high hit rate risks saturating the digital readout circuit, potentially causing data packet loss. 
However, $1.2\times 10^{8}$ hits/second is the nominal continuously sustained allowed hit rate, which is primarily constrained by the digital transfer bandwidth.
In our experiment, we intentionally employed an extremely short exposure time of \SI{50}{\micro s} with a significant interval of 1~s between subsequent exposures. This strategy allowed the Timepix3 to offload internal buffers between two consecutive exposures, enabling us to surpass the nominal maximum hit rate of $1.2 \times 10^{8}$~hits/second. 
Nevertheless, slight overloading persisted. As a consequence, the histogram of the cumulative ToT values across the pixels engaged in a single electron hit is deformed, as illustrated in Fig.~\ref{fig: Fig3_SuppMat}, making it challenging to determine the correct ToT parameter from it.
Therefore, in such scenarios, we chose to refrain from using the ToT parameter as a threshold in the clustering algorithm.
To overcome this challenge, we introduced an additional step in the clustering algorithm. This step involves regrouping events occurring within a distance of fewer than 8 pixels and within a time window of approximately 2~ps. In practice, this step resulted in considering simultaneous counts within 2~ps that are closer than 8 pixels as a single event. 
Before the introduction of this second clustering step, the algorithm had identified a total of 1626 such events, constituting 1.3$\%$ of the total number of identified two-electron events. 
Conversely, the expected number of two-electron events with a separation less than 8 pixels occurring within a time window $\Delta T= 2$~ps is estimated to be $N_2 = N_\mathrm{e}\; \frac{1}{\Delta T} \int_{0}^{\Delta T} \lambda \, e^{-\lambda} \, d\tau = 55$, where $\lambda$ is the average number of electrons in 8 pixels and $ N_\mathrm{e}$ the total number of measured electrons. 
To adopt a conservative approach, we have derived the estimate for $\lambda$ from the measured average number of electron events within $\Delta t = 2$~ps when only the first clustering step is applied.
Therefore, the expected 55 two-electron events arriving in 2~ps with a separation of fewer than 8 pixels represent, at most, an overestimation of the real number and is still smaller than the identified 1626 events considering only the first clustering step. This substantiates the reliability of the 8-pixel threshold we have chosen.

When analyzing data associated with a hit rate exceeding the nominal maximum, our initial step involves grouping pixels within the same cluster if their activation occurs within the $\Delta$(ToA) upper limit and they are within a distance of $\bar d$. Subsequently, we implement the additional clustering step, which groups events occurring within a 2~ps time window and characterized by a separation of less than 8 pixels. Once the clustering process is concluded, we proceed with the statistical analysis following the methodology outlined in the main paper.

The introduction of the additional clustering step affects the temporal resolution of the method, which is ultimately limited by this 8-pixel threshold corresponding to $240$~fs at the highest magnification used in this study. 
Employing a fast beam blanker would reduce the hit rate on the Timepix3, enabling the use of the ToT as a clustering parameter. This, in turn, would lead to more accurate results and, ultimately, an enhanced temporal resolution, potentially reaching a few femtoseconds per pixel.

\section{The Lissajous Pattern: Microwave-Cavity-Based Electron Beam Deflection and Mathematical Description}
The Lissajous pattern emerges from the superposition of two harmonic motions of the electrons in perpendicular directions, each oscillating at a specific frequency. 
In our RF cavity-based ultrafast transmission electron microscope, these orthogonal harmonic motions are induced by the two transverse on-axis magnetic fields oscillating inside the dual-mode cavity at two different resonance frequencies \cite{Borrelli2023,van2018dual, verhoeven2018high}
\begin{align}
  \label{eq:eq1}
  B_\mathrm{y} &= B_1 \sin{(\omega_1 \, t + \phi_1)}  \\
  B_\mathrm{x} &= B_2 \sin{(\omega_2 \, t + \phi_2)}. \nonumber
\end{align}
Here, $B_1$ and $B_2$ are the magnetic field amplitudes, $\phi_1$ and $\phi_2$ the microwave phases at $t = 0$, while $\omega_1 = 2\pi\times f_1$~GHz and $\omega_2= 2\pi\times f_2$~GHz, with $f_1 = 3.000$~GHz and $f_2 = 3.075$~GHz are the cavity's resonance frequencies. 
An electron initially traveling along the magnetic axis of the cavity is periodically deflected off-axis by the two magnetic fields in Eq.~\ref{eq:eq1}. The result is the emergence of a Lissajous pattern in a plane transverse to the cavity axis.
The microwave signals driving the two cavity modes are derived from the 40$^{\mathrm{th}}$ and 41$^{\mathrm{st}}$ harmonics of the same 75~MHz driving signal. 
As a result, one side of the Lissajous pattern consists of 40 nodes and is traced in 167~ps, while the other comprises 41 nodes and is traced in 162~ps.

Mathematically, the Lissajous pattern is described by a vectorial function $ L: t \mapsto (x, y)$, the {\it Lissajous function}, that describes the transverse coordinate of any electron in the beam.
This function can be determined by solving the equation of motion for a generic electron in the beam inside the cavity. In the following treatment, we will consider the longitudinal velocity of the electrons to be constant during the motion ($z' = \frac{\delta v_\mathrm{z}}{ v_\mathrm{z}} \ll 1 $) and neglect the small transverse deviation $x' = \frac{ v_\mathrm{x}}{v_\mathrm{z}}\ll 1$ and $y' = \frac{ v_\mathrm{y}}{v_\mathrm{z}}\ll 1$ from the longitudinal axis during the propagation inside the cavity. 
Considering a nominal electron entering the cavity at $t = 0$ and propagating along the cavity axis with constant velocity $\Bar{v} = v_\mathrm{z} \, \hat{z}$, the Lorentz force exerted on this particle while moving through the fields in Eq.~\ref{eq:eq1} can be expressed as
\begin{equation}
\label{eq:eq2}
    \Bar{F} = q_\mathrm{e} \Bar{v} \times \Bar{B} = q_\mathrm{e}
    \begin{pmatrix}
       -  v_\mathrm{z} \, B_1 \, \sin{(\omega_1 \, t + \phi_1)}  \\
         v_\mathrm{z} \, B_2 \, \sin{(\omega_2 \, t + \phi_2)} \\
        0
    \end{pmatrix}
    = \frac{d\Bar{p}_\mathrm{e}}{dt},
\end{equation}
for $0 \leq t \leq t_{\mathrm{out}}$, with $t_{\mathrm{out}}$ being the time instant at which the electron exits the cavity.
Here, $q_\mathrm{e}$ is the electron charge, $\Bar{p}_\mathrm{e} = m \gamma \bar{v}$ the electron momentum, and $\gamma$ the electron Lorentz factor. This equation is valid under the assumption of a top-hat profile of the on-axis magnetic field amplitudes $B_{1,2} (z) = B_{1,2}$. 
Integrating twice Eq.~\ref{eq:eq2} gives 
\begin{align}
    x(t) &= \frac{q_\mathrm{e} v_\mathrm{z} B_1}{m\gamma \omega_1^2}\left[ \sin(\omega_1 t + \phi_1) - \sin(\phi_1) - \omega_1 t \cos(\phi_1)\right] \\
     y(t) &= - \frac{q_\mathrm{e} v_\mathrm{z} B_2}{m\gamma \omega_2^2}\left[ \sin(\omega_2 t + \phi_2) - \sin(\phi_2) - \omega_2 t \cos(\phi_2)\right]. \nonumber
\end{align}
The transverse coordinates of the nominal electron at the cavity exit are
\begin{align}
    x(t_{\mathrm{out}}) &= \frac{q_\mathrm{e} v_\mathrm{z} B_1}{m\gamma \omega_1^2}\left[ \sin\left(\omega_1 \frac{L_{\mathrm{cav}}}{v_\mathrm{z}} + \phi_1\right) - \sin(\phi_1) - \omega_1 \frac{L_{\mathrm{cav}}}{v_\mathrm{z}} \cos(\phi_1)\right] \\
     y(t_{\mathrm{out}}) &= - \frac{q_\mathrm{e} v_\mathrm{z} B_2}{m\gamma \omega_2^2}\left[ \sin\left(\omega_2 \frac{L_{\mathrm{cav}}}{v_\mathrm{z}} + \phi_2\right) - \sin(\phi_2) - \omega_2 \frac{L_{\mathrm{cav}}}{v_\mathrm{z}} \cos(\phi_2)\right], \nonumber
\end{align}
where $t_{\mathrm{out}} = \frac{L_{\mathrm{cav}}}{v_\mathrm{z}}$ is the transit time of the nominal electron through the cavity, with $L_{\mathrm{cav}}$ being the cavity length. Additionally, $\phi_1$ and $\phi_2$ are the phases of the magnetic fields seen by the nominal electron when it enters the cavity at $t=0$. Any other electron in the beam entering the cavity at a generic time $t$ will experience initial phases $\phi_{1,2} = \phi_{1,2} + \omega_{1,2}\;t$. 
Therefore, the transverse coordinates of a generic electron at the cavity exit are
\begin{align}
\label{eq:eq5}
    x(t) &= A_1 \left[ \sin\left(\omega_1 t + \phi_1+ \omega_1 \frac{L_{\mathrm{cav}}}{v_\mathrm{z}} \right) - \sin(\omega_1 t + \phi_1) - \omega_1 \frac{L_{\mathrm{cav}}}{v_\mathrm{z}} \cos(\omega_1 t +\phi_1)\right] \\
     y(t) &= - A_2 \left[ \sin\left(\omega_2 t +\phi_2 + \omega_2 \frac{L_{\mathrm{cav}}}{v_\mathrm{z}}\right) - \sin(\omega_2 t +\phi_2) - \omega_2 \frac{L_{\mathrm{cav}}}{v_\mathrm{z}} \cos(\omega_2 t +\phi_2)\right], \nonumber
\end{align} 
where $A_1 = \frac{q_\mathrm{e} v_\mathrm{z} B_1}{m\gamma \omega_1^2}$ and $A_2 =  \frac{q_\mathrm{e} v_\mathrm{z} B_2}{m\gamma \omega_2^2}$.
This expression holds true assuming that all the electrons in the beam travel along the cavity axis with constant velocity $v_\mathrm{z}$.
The vectorial function $ L: t \mapsto (x, y)$, with $x$ and $y$ defined as in Eq.~\ref{eq:eq5} is the sought Lissajous function. 
The spacing of the inner lines within the Lissajous figure depends on the phase difference between the two cavity modes $\phi_1 - \phi_2$, which thus determines the appearance of the Lissajous figure. 
The $x$ ($y$) size of the Lissajous pattern can be controlled by adjusting the RF power $W_1$ ($W_2$) feeding the cavity. 

\section{Data Analysis Strategy: Lissajous Pattern Fitting and Electron Correlation Analysis }
This section provides a comprehensive overview of the sequential steps involved in the data analysis process employed in this study. 
In the preliminary phase, we employed a fitting algorithm to fit a tailored version of the Lissajous function $L(t)$ defined in Eq.~\ref{eq:eq5} to a measured pattern. This adapted function was specifically designed to describe real-world measured patterns, accounting for rotations due to the microscope's imaging lens, lack of sharp focus, and potential minor astigmatism leading to a slight shearing of the pattern arising from non-uniform stretching of the image.
The function used to describe these rotated and distorted Lissajous patterns can be expressed as $L_{\mathrm{FIT}}:~t~\mapsto~[x_{\mathrm{FIT}} (A_1, \phi, A_2, x_0, y_0, \theta, \sigma, \kappa), y_{\mathrm{FIT}}(A_1, \phi, A_2, x_0, y_0, \theta, \sigma, \kappa)]$, where

\begin{align}
\label{eq:fitting_function}
    x_{\mathrm{FIT}}(t) &= \cos(\theta) \; x(t) - \sin(\theta) \; y'(t) + x_0\\
    y_{\mathrm{FIT}}(t) &= \sin(\theta) \; x(t) + \cos(\theta) \; y'(t) + y_0, \nonumber
\end{align}

with $y'(t) = x(t) \; \sigma + y(t)$.
Here, $x$ and $y$ are defined similarly to Eq.~\ref{eq:eq5} as
 \begin{align}
    x(t) &= A_1\left[ \sin\left(\omega_1 t + \phi+ \omega_1 \frac{L_{\mathrm{cav}}}{v_\mathrm{z}} \right) - \sin(\omega_1 t + \phi) - \omega_1 \frac{L_{\mathrm{cav}}}{v_\mathrm{z}} \cos(\omega_1 t +\phi)\right] \\
     y(t) &= A_2 \left[ \sin\left(\omega_2 t + \omega_2 \frac{L_{\mathrm{cav}}}{v_\mathrm{z}}\right) - \sin(\omega_2 t ) - \omega_2 \frac{L_{\mathrm{cav}}}{v_\mathrm{z}} \cos(\omega_2 t)\right]. \nonumber
\end{align}
Considering the intricate nature of the Lissajous pattern, employing image fitting instead of function fitting results to be more efficient.
A Lissajous pattern captured by the Timepix3 detector is a gridded image comprising $514\times514$ individual pixels. 
To adequately represent the pixellated character of the measured patterns, we, therefore, computed the function $L_{\mathrm{FIT}}$ over an identical grid, aggregating values from specific locations. This computation transforms $L_{\mathrm{FIT}}$ into a fitting image.
Furthermore, we applied a Gaussian blur filter, characterized by a blurring intensity parameter $\kappa$, to this Lissajous fitting image to reproduce the finite beam spot size in the measured image. 
The fitting image $L_{\mathrm{FIT}}$ depends on 8 free parameters, encompassing 
the pattern's rotation angle $\theta$, the coordinates of the pattern center $(x_0, y_0)$, the shearing parameter $\sigma$, the blurring parameter $\kappa$, and the normalized amplitude factors $A_1$ and $A_2$ determining the length of the Lissajous sides. As previously discussed, the phases in the Lissajous function definition (Eq.~\ref{eq:eq5}) govern the spacing of the pattern's inner lines, a feature solely reliant on the phase difference $\phi_1 - \phi_2$. Hence, we have only included one phase parameter $\phi$ in the fitting function definition.
Additional parameters within the argument of the fitting function hold predefined values, such as the cavity resonance frequencies $\omega_1 = 2\pi \times3.000$~GHz and $\omega_1 = 2\pi \times3.075$~GHz, the cavity length $L_{\mathrm{cav}} = 16.67$~mm, and the electron velocity $v_{\mathrm{z}} = 0.6953\times c$, where $c$ is the speed of light.

Given the considerable number of free parameters involved in $L_{\mathrm{FIT}}$, we adopted an iterative strategy for the fitting procedure.
Initially, $L_{\mathrm{FIT}}$ was fitted to the measured pattern using all the available free parameters except for the phase $\phi$ to minimize the distance between the two images. This initial step aims to match their shapes. At this stage, we assigned a fixed value to $\phi$ producing a pattern with inner lines roughly resembling those in the measured image.
From the measured coordinates of the Lissajous figure corners $(x_\mathrm{i},y_\mathrm{i})$ with i = 1,2,3,4, we calculated the fitting starting values of the rotation angle, amplitudes, and center as follows:
\begin{align*}
    &(x_0, y_0) = \left(\frac{1}{4} \sum_{i=1}^4{x_\mathrm{i}} , \frac{1}{4} \sum_{i=1}^4{y_\mathrm{i}}\right) \\
    &\theta = \arccos \left({- \frac{y_2 - y_3}{\sqrt{(x_2-x_1)^2 + (y_2-y_1)^2}}}\right) \\
    &A_1 = \frac{1}{2}  \left(\sqrt{(x_2-x_1)^2 + (y_2-y_1)^2} + \sqrt{(x_3-x_4)^2 + (y_3-y_4)^2} \right) \\
    &A_2 = \frac{1}{2}  \left(\sqrt{(x_2-x_3)^2 + (y_2-y_3)^2} + \sqrt{(x_1-x_4)^2 + (y_1-y_4)^2} \right).
\end{align*}
We employed a least-squares fitting algorithm, which minimizes the sum of square residuals (SSR) to determine the best-fitting parameters for $L_{\mathrm{FIT}}$. 

In the second phase of the fitting procedure, the free parameters considered in the prior step were held constant at the determined best-fitting values, while the primary objective shifted towards optimizing $\phi$ to minimize the SSR.
This optimization requires specifying a restricted range wherein the search for the optimal phase is confined due to the existence of numerous phase values leading to locally minimized differences between the fitting function and the measured pattern.
However, constraining the phase range alone is insufficient to ensure convergence. For the subsequent correlation analysis, we exclusively utilize magnified Lissajous patterns, which ensure higher temporal resolution. 
Through careful microscope alignment, we achieved magnification around a specific zooming point without introducing any rotation, translation, or deformations to the Lissajous pattern. Consequently, the same Lissajous function that describes the entire figure also describes the magnified pattern when subject to the same degree of zoom. To facilitate phase optimization, we thus utilized the magnified Lissajous pattern, consisting of only a few lines and thereby reducing the population of local minima.  
More precisely, we appropriately adjusted the scaling of the fitting image evaluated in the best-fitting parameters obtained from the initial fitting step to match the magnified pattern. We then fitted this scaled image to the magnified pattern, considering $\phi$ as the sole free parameter. 

The first and second fitting steps were reiterated using the most current set of best-fit parameters in each iteration until convergence was achieved with SSR below $10^{-6}$.

\section{Beam Coherence}

The transverse coherence length $l_{\mathrm{x,y}}$ of an electron beam is defined by the expression
\begin{equation}
   l_\mathrm{{x,y}} = \frac{\hbar}{ \sigma_{\mathrm{p_{x,y}}}}.
\label{eq:transverse_coherence_length}
\end{equation}
Here, $\sigma_{\mathrm{p_{x,y}}}$ is the (rms) transverse momentum spread, determined as  $\sigma_{\mathrm{p_{x,y}}} = \frac{ m c \, \epsilon_{\mathrm{x,y}}}{\sigma_{\mathrm{x,y}}}$, with 
$\sigma_{\mathrm{x,y}}$ and $\epsilon_{\mathrm{x,y}}$ being the transverse (rms) beam spot size and emittance, respectively.
The (rms) transverse sizes and normalized emittances of the electron beam emitted by the Schottky field emission gun of the TU/e UTEM, were measured to be $\sigma_{\mathrm{x,y}} \approx 1$~nm and $\epsilon_{\mathrm{x,y}} \approx 2$~pm~rad, at a peak current $I = 0.4$~nA \cite{van2018dual}.
Correspondingly, the calculated transverse coherence lengths from Eq.~(\ref{eq:transverse_coherence_length}) are $l_{\mathrm{x,y}} \approx 200$~pm, resulting in a degree of degeneracy $l_{\mathrm{x,y}}/ \sigma_{\mathrm{x,y}}= 0.2$. 
Therefore, the peak (rms) brightness of the electron beam emitted by our Schottky FEG is
\begin{equation}
   B_{\mathrm{n}} = \frac{q}{ m c^2} \frac{I}{4 \; \pi^2\; \epsilon_{\mathrm{x}}\epsilon_{\mathrm{y}}} = 7 \times 10^6~\frac{\mathrm{A}}{\mathrm{m}^2\mathrm{srV}},
\end{equation}
resulting in $\epsilon_{\mathrm{x}} \sim \epsilon_{\mathrm{y}} \approx 21$~pm~rad at a peak current $I~=~46.2$~nA.
Consequently, the degree of transverse coherence for the electron beam employed in the experiment discussed in this paper is $l_{\mathrm{x,y}}/ \sigma_{\mathrm{x,y}}= 0.02$. 

The width of the lines in the Lissajous pattern directly correlates with the transverse (FWHM) spot size of the electron beam. 
In the magnified pattern shown in Fig.~3c of the main paper, used for the statistical measurement, the line width measures 0.56~mm, corresponding to a transverse (rms) spot size $\sigma_{\mathrm{x,y}} \approx$~\SI{0.12}{\micro\m}.
Therefore, the transverse coherence length of the electron beam used in the measurements included in this paper is $l_{\mathrm{x,y}} \approx 2$~nm.

The coherence time $\tau_\mathrm{c}$ of an electron beam can be directly computed using the Heisenberg energy-time uncertainty principle
\begin{equation}
   \tau_\mathrm{c} \geq \frac{\hbar}{2\sigma_\mathrm{E}},
\end{equation}
where $\sigma_\mathrm{E}$ is the (rms) energy spread of the beam. 
For a Schottky field emission gun (FEG) characterized by an (rms) energy spread $\sigma_\mathrm{E} \leq 0.5$~eV, the (rms) coherence time is $\tau_\mathrm{c} \geq 0.8$~fs.

The beam degeneracy, defined as the number of electrons within a coherence volume $V_\mathrm{c}$, is determined from the number $n$ of electrons in any given beam volume $V_\mathrm{n}$ according to the relation 
\begin{equation}
 \delta = \frac{n}{2} \frac{V_\mathrm{c}}{V_\mathrm{n}} = \frac{1}{2} \frac{V_\mathrm{c}}{V_\mathrm{1}}. 
\end{equation}
Here, the factor $1/2$ accounts for the spin degeneracy in an unpolarized electron beam. 
Assuming that a Schottky FEG emits an electron beam with constant velocity and cylindrical symmetry \cite{bronsgeest2009physics}, the beam degeneracy can be expressed as
\begin{equation}
    \delta =  \frac{1}{2} \frac{A_\mathrm{c}}{A} \frac{\tau_\mathrm{{c}}}{\sigma_\mathrm{t}}, 
\end{equation}
where $A = \sigma_{\mathrm{x}} \cdot \sigma_{\mathrm{y}}$ is the beam cross-section, $\sigma_\mathrm{t}$ is the (rms) time duration of a single electron volume length, $A_\mathrm{c} = l_{\mathrm{x}} \cdot l_{\mathrm{y}}$ is the transverse coherence area, and $\tau_\mathrm{{c}}$ is the coherence time. 
The beam degeneracy thus depends on both the degrees of longitudinal $\frac{\sigma_\mathrm{{\tau_{c}}}}{T}$ and transverse $\frac{l_{\mathrm{x,y}}} {\sigma_{\mathrm{x,y}}}$ coherence. Using  $\frac{A_\mathrm{c}}{A} = \frac{l_{\mathrm{x}}} {\sigma_{\mathrm{x}}} \cdot \frac{l_{\mathrm{y}}}{\sigma_{\mathrm{y}}} \approx 4\times10^{-4}$ and $\frac{\sigma_\mathrm{{\tau_{c}}}}{\sigma_\mathrm{t}} \approx 5 \times 10^{-4}$ ($\sigma_\mathrm{t} \approx 1.5$~ps) we determine that the degeneracy of the 46.2~nA electron beam used in this study is $\delta \approx  2 \times 10^{-7}$.

\section{Simulative study of the electron emission from a Schottky field emission gun in General Particle Tracer}

To investigate the impact of the Coulomb repulsion on the suppression of higher electron number events within the time windows analyzed in this study, we developed a General Particle Tracer (GPT) simulation \cite{Bas2022,van20053d,poplau2004multigrid} that reproduces the emission of a 46.2~nA electron beam from a simplified yet realistic model of the Schottky field emission of our UTEM. 
The simulation is based on the Boundary Element Method (BEM) solver in GPT, which enables solving electrostatic fields in complex geometries under the assumption of perfectly conducting surfaces at fixed potentials.

\begin{figure}[h]
    \centering
    \includegraphics[width=0.98\textwidth]{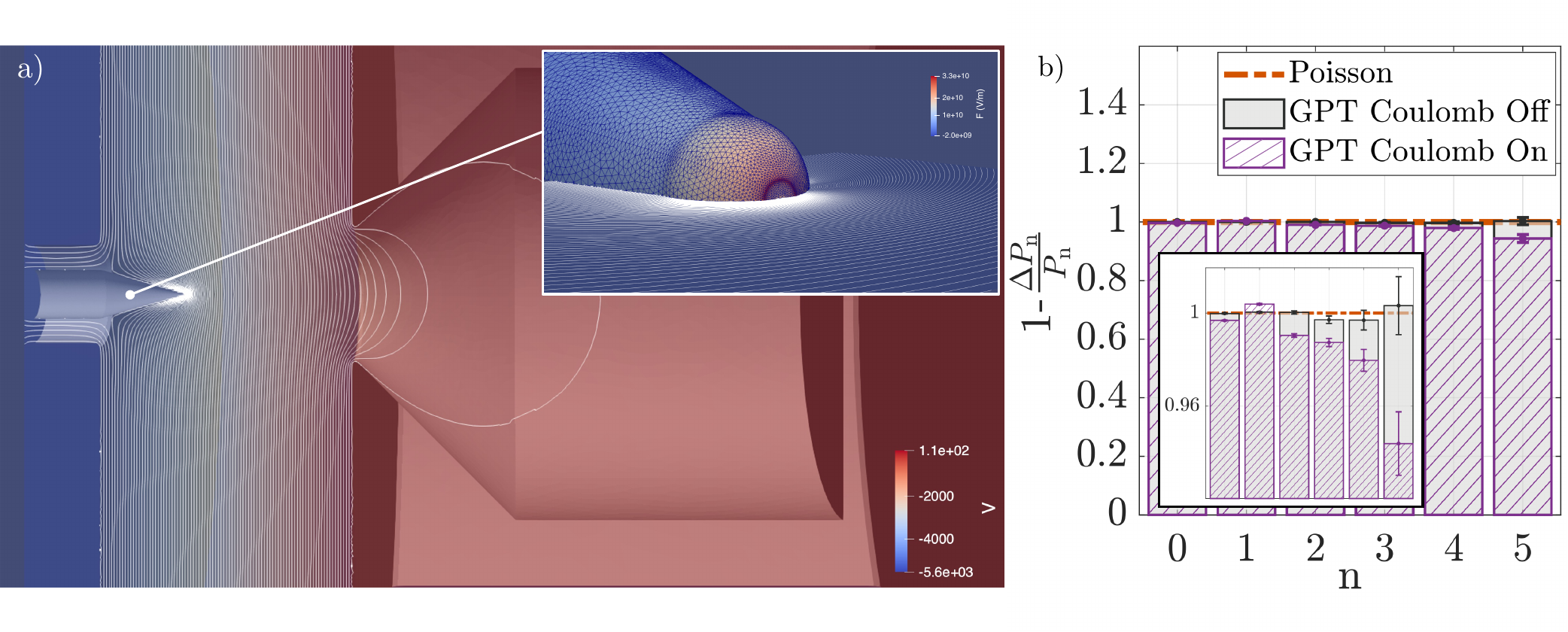}
  \caption{(a) Cut-through view of the simulated tip and gun geometry in GPT, with color-coding on electric potential intensity. The inset shows a zoomed-in perspective of the tip, color-coded on the electric field strength. The contour plots show the equipotential lines. (b) Relative deviation from the expected Poisson distribution $\Tilde{P}_\mathrm{n}$, with $\langle n \rangle = 0.57$, of the simulated probability of 0, 1, 2, 3, 4, and 5 simultaneous electron events within a  2~ps time window. The comparison is presented between two scenarios, one accounting for Coulomb interaction and the other without. The inset provides a closer look at the deviations. \label{fig:GPT_gun}}
\end{figure}

We employ the \textit{BEMdraw} package of the BEM solver to create a realistic three-dimensional model of the Schottky FEG in our TEM. 
Figure \ref{fig:GPT_gun}a presents a cut-through view of the simulated gun geometry, with color-coding based on the electric potential intensity. The inset provides a zoomed-in view of the emitter tip, with color-coding on the local electric field strength. The contour plots in the main image and the inset illustrate the equipotential lines of the electromagnetic field.

The emitter of our gun consists of a $\approx 1$~mm long single-crystalline tungsten wire with a diameter of $\approx$~\SI{125}{\micro\m}, one end of which is etched down to a tip with a diameter of \SI{0.9}{\micro\m}. 

A distinctive feature of a Schottky emitter is the presence of crystallographic planes with ${100}$ orientation characterized by a reduced work function. The reduced work function along these planes, compared to other crystallographic orientations, is attributed to the presence of a ZrOx reservoir located halfway along the wire emitter, which diffuses up to these planes at high operating temperatures. The lower work function results in electron emission predominantly from these planes.
The primary emission plane, responsible for most of the emission process, is the facet on the tip end, which, in our case, has a diameter of $\approx300$nm. To streamline the simulation process, we model only this end facet, neglecting the emission from the other ${100}$ planes.
Electron emission is induced by heating the source and applying an electric field. The standard operating temperature of our Schottky emitter is 1800~K and the exciting current through the emitter is approximately 2.3~A. 

The electric field is applied by negatively biasing the emitter relative to the extractor (leftmost element in Fig.~\ref{fig:GPT_gun}a), which we simulate as a metallic plate with a central aperture of $\approx$~\SI{380}{\micro\m} at a distance of $0.5$~mm from the end facet of the tip.
As in all standard configurations, our Schottky gun includes a Wehnelt suppressor, a metal cup (rightmost element in Fig.~\ref{fig:GPT_gun}a) with a central aperture of $\approx$~\SI{250}{\micro\m} through which the emitter tip protrudes for $0.27$~mm. The suppressor is negatively biased relative to the emitter to suppress unwanted electron emission from the part of the emitter inside the cap.
In our simulation, the emitter is grounded, and the suppressor and extractor are maintained at operating voltages of $- 0.3$~kV and 5~kV, respectively. The extraction current from our gun is \SI{32}{\micro\A}.

After reproducing the gun geometry with \textit{BEMdraw}, we utilize the \textit{BEMmesh} package to create a refined 3D surface mesh. Subsequently, we employ the \textit{BEMsolve} program to solve the boundary element method equations for each meshed electrode modeled.
The obtained results are then imported into GPT using the \textit{BEMcharges} element, which scales the \textit{BEMsolve} solution to match the actual potentials on the electrodes, enabling the calculation of realistic electromagnetic fields in the gun region and beyond. 


Using GPT, we trace the propagation of an electron beam with a current of $I = 46.2$~nA, equivalent to the current used in our experiment, through the computed electromagnetic fields, and then image it onto a screen located 0.6~mm away from the emitter tip.
In the actual experimental setup, the microscope's condenser system is responsible for shaping and sizing the beam. 
However, in our simplified model, we focused on simulating the electron propagation in the immediate vicinity of the end facet of the tip, without including the condenser system. 
As we do not simulate the condenser system, we reproduce the beam used during the experiment by assessing the effective emission surface of the tip resulting in a 46.2~nA beam current. To achieve this, we utilize the expression for the emission current from a Schottky emitter \cite{bronsgeest2009physics} given by:
\begin{equation}
j(F,T) = \frac{qm}{2\pi^2\hbar^3} \; (k_\mathrm{B}\,T)^2 \; e^{-\frac{\phi-\sqrt{\frac{q^3F}{4\phi \varepsilon_0}}}{k_\mathrm{B}\,T}}.
\end{equation}
Here, $q$ and $m$ denote the electron charge and mass, $k_\mathrm{B}$ is the Boltzmann constant, and $\varepsilon_0$ is the vacuum permittivity. $T = 1800$~K is the emitter operating temperature, and $\phi = 2.93$V the work function of our emitter. 
The longitudinal electric field is approximately constant within 1-2~nm from the center of the end facet surface, with a strength of about $F\sim 1$ GV/m.
At the operating parameters of our gun, the resulting current density is $j(F,T) \approx 5.6 \times 10^7$ A/m$^2$, and the corresponding tip radius to achieve a current of 46.2~nA is $r\approx 16$~nm.

In our simulation, we model the emission of an electron beam with a 16~nm radius starting from a distance of 2~nm from the end facet surface, assuming a random uniform distribution. The simulation neglects the angular spread of the emitted beam.
To initiate the electron beam, we assigned random starting times following a uniform distribution between $t/2$ and $t$, where $t = \frac{n_\mathrm{e} q_\mathrm{e}}{I}$. Here, $n_\mathrm{e} = 10000$ represents the number of simulated particles, and $q_\mathrm{e}$ is the electron charge.
We trace the simulated electron beam as it travels through the electromagnetic fields of the gun up to the screen, exploring two different scenarios. In one simulation, we account for all pairwise stochastic Coulomb interactions among electrons in the beam, while in another simulation we exclude them.

The primary outcome of the simulation is a time vector containing the arrival times of the $n_\mathrm{e}$ simulated electrons on the screen. Employing the same statistical analysis as detailed in the paper, we assessed the probability distribution $P_\mathrm{n}$ of the number $n$ of simultaneous electrons arriving on the screen within a time window $\Delta t = 2$~ps, for both considered scenarios. 
To enhance statistical significance, the simulation was iteratively conducted until reaching a total number of simulated electrons approximately equal to 8 million. 

Figure \ref{fig:GPT_gun} illustrates the relative deviations of $P_\mathrm{n}$ for $n = 0, 1, 2, 3, 4, 5$ from the expected Poisson distribution $\Tilde{P}_\mathrm{n} = \frac{\langle n \rangle^n e^{-\langle n \rangle}}{n!}$ in both simulated scenarios. Here, $\langle n \rangle$ represents the average number of electrons in the time window $\Delta t$ as obtained from the simulation.
The relative deviations are computed as $1 - \frac{\Delta P_\mathrm{n}}{\Tilde{P}_\mathrm{n}}$, where $\Delta P_\mathrm{n} = P_\mathrm{n} - \Tilde{P}_\mathrm{n}$. 
The displayed error bars represent the statistical uncertainties associated with observing $N_\mathrm{n}$ simultaneous $n$ electron events, with $n$ ranging from 0 to 5. These errors are determined by the expression $ \frac{\sqrt{N_\mathrm{n} \left(1- \frac{N_\mathrm{n}}{N_{\mathrm{exp}}} \right)}}{\widetilde{P}_\mathrm{n} \;  N_{\mathrm{exp}}}$, with $N_{\mathrm{exp}}$ denoting the total number of measurements. 
In the absence of Coulomb interaction, we observe a Poisson distribution within the considered 2~ps time window. However, upon introducing Coulomb interaction, the distribution shifts to sub-Poissonian when evaluated within the same 2~ps time window.

Our simulation reveals a sub-Poissonian behavior arising from stochastic electron-electron Coulomb interaction in the beam emitted from a realistic Schottky FEG. The anti-bunching is observed within the same time window employed in the experiment detailed in this paper.
However, these are preliminary results that do not permit definitive conclusions. 

\nocite{*}

\bibliography{2Supplementary}